# Quantum Measurement Without Collapse or Many Worlds: The Branched Hilbert Subspace Interpretation

Xing M. Wang[1]

## Abstract

We propose the Branched Hilbert Subspace Interpretation (BHSI) as an alternative view of quantum measurement. BHSI explains measurement as a unitary branching of the local Hilbert space (LHS) into decoherent, independent, and unitarily evolving subspaces. It also updates observer states (through their equipment) by causally engaging and disengaging operators. Unlike the Copenhagen Interpretation (CI), BHSI avoids wave function collapse while maintaining the Born rule through branch weights linked to the initial system state. Unlike the Many-Worlds Interpretation (MWI), BHSI avoids parallel worlds by entangling branches with the local environment within a single world. We compare BHSI's features with those of CI, MWI, and Bohmian Mechanics (BM). We explore its implications for the double-slit experiment, Bell tests, Wigner's friend, black hole radiation, and the delayed-choice quantum eraser. We analyze quantum teleportation and show that locally controlled decoherence and recoherence processes (CDRP) can be observed. Specifically, we propose experiments using modern Stern-Gerlach interferometers (SGIs) to visualize CDRP, measure branch weights that encode the Born rule, and predict the electromagnetic (EM) phase shift arising from the independent unitary evolution of decoherent branches. BHSI thus offers a minimalist alternative to interpretations based on collapse or on many worlds.

**Keywords:** quantum foundations; measurement problem; local unitary branching; Born rule; local Hilbert space.

## 1. Introduction

The interpretation of quantum mechanics (QM) has been a subject of debate since its inception in the 1920s. The theory's mathematical formalism, such as unitary evolution, superposition, and entanglement, yields strikingly non-classical predictions, yet its physical meaning remains contested. The Copenhagen Interpretation (CI; Bohr, Heisenberg, Born, Pauli, 1920s-1950s [1-3]) provides a mathematically simple framework that aligns with laboratory observations. However, it faces criticism for its ill-defined wave-function collapse, the straightforward postulation of the Born rule [4], the cornerstone of QM's probabilistic predictions, and the subjective boundary between the quantum and classical regimes. The Many-Worlds Interpretation (MWI; Everett, DeWitt, Deutsch, Wallace; 1957-present, [5-7]) addresses the measurement problem by postulating that all possible quantum measurement outcomes occur in separate, non-interacting branches of reality (each branch is a world with a copy of the observer), thereby offering a compelling solution by eliminating wavefunction collapse. Still, it faces significant challenges, including ontological excess, the lack of a convincing explanation for the Born rule, and the preferred-basis issue [8-11]. Bohmian Mechanics (BM, Bohm, Bell, Goldstein; 1952-present; [12-14]), also known as the de Broglie-Bohm pilot-wave theory, resolves the wave collapse issue of CI within a single world, but it relies on hidden variables (actual particle positions), and its explicit nonlocality structure may conflict with relativity.

[1] Sherman Visual Lab, Sunnyvale, CA 94085, USA; xmwang@shermanlab.com; ORCID:0000-0001-8673-925X

We propose an alternative approach: the Branched Hilbert Subspace Interpretation (BHSI), in which measurement divides the local Hilbert space (LHS) into multiple branches rather than splitting the universe into parallel worlds within the global Hilbert space. Since each possible outcome exists and evolves within a single branch, no wave function collapses. The observer's state is updated in a relational and causal manner, yielding a single outcome per observation. The Born rule [4] can be realized through the branch weight (probability) linked to its amplitude in the initial system state, on the basis chosen by the observer. With only one observer in a single world, it avoids the ontological challenge of explaining probability in the MWI.

In Section 2, we formalize the mathematical framework of BHSI by defining unitary operators for branching, engaging, and disengaging. We explain how these operators decohere subspaces and update the observer's state. We demonstrate how the Born rule applies to measurements and what a local observer would observe. In Section 3, we compare the features of CI, MWI, and BM with those of BHSI. In Section 4, we contrast BHSI with CI and MWI by examining their implications for interference (double-slit experiment [15, 16]), nonlocality (Bell tests [17, 18]), causal dominance (Wigner's friend) [5, 20, 21]), black hole radiation with the No-Hiding Theory [22, 23], and the delayed choice quantum eraser [24, 25]. In Section 5.1, we analyze the nature of branching by comparing the environmental scale of quantum decoherence [26-28] in MWI and BHSI (maximum versus minimum). In Section 5.2, we describe how BHSI's locally controlled decoherence-recoherence process (CDRP) has already been observed in quantum teleportation [29]. In Section 5.3, we propose experiments to directly visualize the CDRP, measure the weights carried by the branches (preserving the Born rule), and detect the electromagnetic (EM) phase shifts caused by the independent unitary evolution of decoherent branches using modern Stern–Gerlach interferometers (SGIs) [30, 31]. We briefly compare BHSI with other Single-World Interpretations (SWIs) in Appendix A.

## 2. Mathematical Framework.

In this section, we present the fundamental concepts of BHSI: branching LHS, updating (engaging and disengaging) the observer's state, and the Born rule.

### 2.1. The Branching, Engaging, and Disengaging Operators

Assume the observer chooses to measure an observable $\hat{G}$. The following linear combination on the $G$-basis describes the initial quantum state ([2, p.29]):

$$|\Psi\rangle = \sum_{i=1}^{D} c_i\,|g_i\rangle, \quad \hat{G}\,|g_i\rangle = g_i\,|g_i\rangle, \quad \langle g_i\,|\,g_j\rangle = \delta_{i,j}, \quad \sum_{i=1}^{D}|c_i|^2 = 1, \quad \prod_{i=1}^{D}|c_i| \neq 0 \tag{1}$$

The initial Hilbert space is $D$-dimensional, corresponding to the $D$ possible measurement outcomes, each with a non-zero probability. The *branching operator* $\hat{B}$ is a unitary operator that splits the $D$-dimensional Hilbert space $\mathcal{H}^D$ into $D$ branches:

$$\hat{B} \equiv \sum_{k=1}^{D}|g_{B;k}\rangle\langle g_k|, \quad \hat{B}^{\dagger}\hat{B} = I, \quad \hat{B}\hat{B}^{\dagger} = I_B, \ \langle g_k\,|\,\Psi\rangle = \langle g_k\,|\,BB^{\dagger}\,|\,\Psi\rangle \equiv \langle g_{B,k}\,|\,\Psi_B\rangle = c_k \tag{2}$$

$$\hat{B}(|\Psi\rangle\otimes|E\rangle_L) = |\Psi_B\rangle = \sum_{k=1}^{D} c_k\,|g_k\rangle\,|E_k\rangle_L = \sum_{k=1}^{D} c_k\,|g_{B;k}\rangle, \quad |g_{B;k}\rangle \equiv |g_k\rangle\,|E_k\rangle_L$$
$$\hat{B}\left(\mathcal{H}_S\otimes\mathcal{H}_L\right) = \bigoplus_{k=1}^{D}\mathcal{H}_{S,k}(\text{span } c_k\,|g_{B;k}\rangle), \quad |\langle g_k|\Psi\rangle|^2 = |\langle g_{B;k}|\Psi_B\rangle|^2 = |c_k|^2 \tag{3}$$

Note that the states $|g_{B,k}\rangle$ are locally decoherent, evolving in different branches; the surrounding environment $|E\rangle_L$ is entangled to make them decoherent. Such subspaces are not merely a theoretical construct: *they are observable in experiments* (see Section 5.2-3). The engaging and disengaging operator $\Sigma_\beta \equiv \Gamma_\beta T_\beta \Lambda_\beta$ is a product of three unitary operators.[2] The first operator is the engaging operator $\Lambda_\beta$. It updates the observer's state from |ready⟩ in the environment $\mathcal{H}_E$ to |reads⟩ and entangles the observer's state with the $\beta^{\text{th}}$ subspace. The operator product $\Lambda_\beta \hat{B}$ branches the Hilbert space and randomly engages the observer with the $\beta^{\text{th}}$ subspace:

$$\Sigma_\beta \equiv \Gamma_\beta T_\beta \Lambda_\beta, \quad \Lambda_\beta : |\text{ready}\rangle_o \in \mathcal{H}_E \mapsto |\text{reads } g_\beta\rangle_o \in \mathcal{H}_{S,\beta} \tag{4}$$

$$\Lambda_\beta\hat{B}\text{: } \mathcal{H}_S\otimes\mathcal{H}_L \mapsto \mathcal{H}_B = \bigoplus_{k=1}^{D}\left\{\mathcal{H}_{S,k}(\text{span } c_k\,|g_{B,k}\rangle(|\text{reads } g_\beta\rangle_O)^{\Delta(k,\beta)}\right\}, \quad \beta\in\{1.2.\cdots D\} \tag{5}$$

To simplify the expression, we have used the following notation:

$$\Delta(k,\beta) = \delta_{k,\beta} = \begin{cases}1, \text{ if } k=\beta\\ 0, \text{ if } k\neq\beta\end{cases} \text{ (discreate case)}, \quad (|\text{reads}\rangle_O)^{\Delta(k,\beta)} = \begin{cases}|\text{reads}\rangle_O, \text{ if } k=\beta\\ 1, \text{ if } k\neq\beta\end{cases} \tag{6}$$

After recording the outcome, operator $T_\beta$ changes the observer's state to |ready⟩, then operator $\Gamma_\beta$ disengages him from the branch, ensuring he is prepared for the next engagement.

$$T_\beta : |\text{reads}\rangle_O \mapsto |\text{ready}\rangle_O; \quad \Gamma_\beta T_\beta : \mathcal{H}_B \mapsto \mathcal{H}_f = \left\{\bigoplus_{k=1}^{D}\mathcal{H}_{S,k}(\text{span } c_k\,|g_{B,k}\rangle\right\}\otimes|\text{ready}\rangle_O \tag{7}$$

Let $U(t)$ be the time evolution operator of the system, which can be relativistic or not:

$$|\Psi\rangle \equiv |\Psi_B(0)\rangle, c_k \equiv c_k(0), \quad U(t)\,|\Psi_B(0)\rangle = |\Psi_B(t)\rangle \tag{8}$$

Each branch evolves unitarily and independently after branching:

$$U(t) = \exp\{(i/\hbar)[\int_0^t \hat{H}(t)dt]\}, \quad U(t)\,|g_{B,k}\rangle = \exp[i\,\Phi_k(t)] \tag{9}$$
$$U(t)\hat{B}\{|\Psi\rangle\} = \sum\nolimits_{k=1}^{D} c_k U(t)\,|g_{B;k}\rangle = \sum\nolimits_{k=1}^{D} c_k e^{i\Phi_k(t)}\,|g_{B;k}\rangle \equiv \sum\nolimits_{k=1}^{D} c_k(t)\,|g_{B;k}\rangle \tag{10}$$

Altogether, a measurement process can be described by a unitary transformation $\hat{M}_\beta$ with $\beta$ as a random choice:

---

[2] They act like the unitary NOT gate, flipping between the observer's states [24, p.233].

$$\hat{M}_\beta \equiv \Sigma_\beta \hat{B} = \Gamma_\beta T_\beta \Lambda_\beta \hat{B}, \quad \hat{M}_\beta^\dagger \hat{M}_\beta = I, \quad \beta \in \{1, 2, \cdots D\} \tag{11}$$

$$\hat{M}_\beta \{|\Psi\rangle \otimes |\text{ready}\rangle_O |E\rangle_L\} = \Gamma_\beta T_\beta \left\{ \sum_{k=1}^{D} c_k |g_{B;k}\rangle \otimes (|\text{reads}\rangle_O)^{\Delta(k,\beta)} \right\} = |\Psi_B\rangle \otimes |\text{ready}\rangle_O \tag{12}$$

Therefore, the *decoherent branches evolve unitarily and independently, with their amplitudes* given by the initial system state after the measurement.

### 2.2. The Measurement Process and the Born Rule in BSHI

The initial Hilbert space is $D$-dimensional, as Eq. (1) describes. We discuss three cases.
**Case 1**: $D$ = 1. The initial normalized state contains only one basis state.

$$|\Psi\rangle = |g_1\rangle \tag{13}$$

Since this reflects the observer's measurement basis, the observer consistently records $g_1$, with $P(g_1)$ =1, by unitarily branching, engaging, and disengaging. Only one branch exists, containing $|g_{B,1}\rangle$ after the measurement. There is no loss of information or gain of entropy.

**Case 2**: $D \geq 2$. Before the observation, the system ($S$), the local environment $|E\rangle_L$, and the state of the observer or the apparatus ($O$) are in the following pre-measurement state:

$$|\Psi_0\rangle = |\Psi\rangle \otimes |\text{ready}\rangle_O |E\rangle_L, \quad |\Psi\rangle = \sum_{k=1}^{D} c_k |g_k\rangle \tag{14}$$

According to Eq. (5), branching the system splits its LHS into $D$ parallel subspaces, each spanning a basis state. The observer engages with one branch, which has an associated weight based on its amplitude in the initial state, thereby realizing the Born rule:

$$\mathcal{H}_S \otimes \mathcal{H}_L \to \bigoplus_{k=1}^{D} \mathcal{H}_{S,k}[\text{span } \mathrm{c}_k |g_{b,k}\rangle(|\text{reads } g_\beta\rangle_o)^{\Delta(k,\beta)}], \quad P(g_\beta) = |c_\beta|^2 = |\langle g_\beta | \Psi\rangle_S|^2 \tag{15}$$

After the measurement, the observer disengages from the branched system state, as illustrated by Eq. (7), and the decoherent branches evolve independently with their initial amplitudes:

$$|\Psi_f\rangle = |\Psi_B\rangle \otimes |\text{ready}\rangle_O = \left\{ \sum_{k=1}^{D} c_k |g_{b;k}\rangle \right\} \otimes |\text{ready}\rangle_O \tag{16}$$

**Case 3**: $D$ = 2. This is a specific example of Case 2: the initial state consists of only two basis states. We aim to use this case to compare step by step with the MWI. Assume that Bob is observing a qubit. Before the measurement, we have:

$$\text{MWI:} \quad |\Psi_0\rangle = (\alpha_0 |0\rangle + \alpha_1 |1\rangle)|B\rangle|E\rangle, \quad |\alpha_0|^2 + |\alpha_1|^2 = 1 \tag{17}$$

$$\text{BHSI:} \quad |\Psi_0\rangle = (\alpha_0 |0\rangle + \alpha_1 |1\rangle)|\text{ready}\rangle_O |E\rangle_L, \quad |\alpha_0|^2 + |\alpha_1|^2 = 1 \tag{18}$$

The equation appears similar. The difference is the scope of the environment involved. After branching, their states have the following forms:

MWI: $|\Psi_f\rangle = \alpha_0 |0\rangle |B_0\rangle |E_0\rangle + \alpha_1 |1\rangle |B_1\rangle |E_1\rangle, \quad \langle B_0 | B_1\rangle \approx 0, \langle E_0 | E_1\rangle \approx 0$ (19)

BHSI: $|\Psi_B\rangle = \sum_{k=0}^{1} \alpha_k |k_B\rangle (|\text{reads } k\rangle_O)^{\Delta(k,\lambda)}, \quad \lambda \in \{0,1\}, \quad P(\lambda) = |\alpha_\lambda|^2$ (20)

Or: $|\Psi_B\rangle = \alpha_0 |0_B\rangle (|\text{reads } 0\rangle_O)^{\Delta(0,\lambda)} + \alpha_1 |1_B\rangle (|\text{reads } 1\rangle_O)^{\Delta(1,\lambda)}, \quad \lambda \in \{0,1\}, \quad P(\lambda) = |\alpha_\lambda|^2$ (21)

In MWI, the original world splits into two independent and never-interacting worlds, each with a real Bob. After reading in BHSI, Bob is disengaged. The final state contains two locally decoherent branches that evolve independently and unitarily with corresponding amplitudes:

BHSI: $|\Psi_B\rangle = \alpha_0 |0_B\rangle + \alpha_1 |1_B\rangle = \alpha_0 |0\rangle |E_0\rangle_L + \alpha_1 |1\rangle |E_1\rangle_L, \quad |\alpha_0|^2 + |\alpha_1|^2 = 1$ (22)

The BHSI borrows the concept of branching from the MWI. However, instead of updating the universal wave function in the global Hilbert space, the BHSI only updates the minimal local environment with Bob reading one of the local branches (see Fig. 1). After the branching, in the MWI, each branch represents a real world with a real Bob, though there is no experimental evidence for this so far. In contrast, for the BHSI, locally decoherent subspaces are observable in quantum teleportation and can be visualized in the Stern–Gerlach interferometer (SGI) experiments (see Section 5.2-3); moreover, there is no real Bob in the LHS but the state of Bob as represented through the engaged part of his apparatus. Since each possible outcome is contained in one branch, which evolves independently and unitarily after measurement, the wave collapse in the CI is avoided without the need for many worlds.

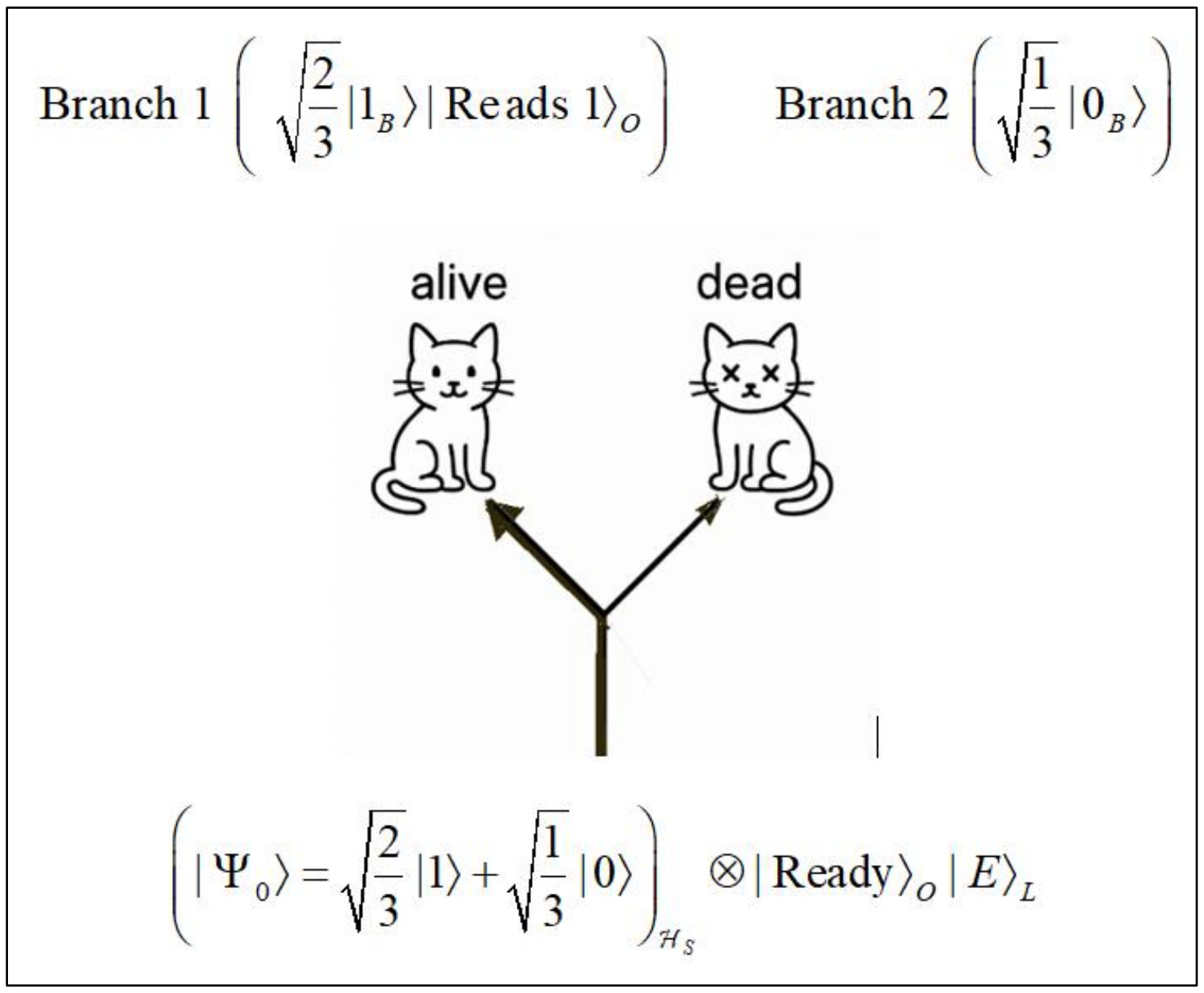

Fig. 1: The Branched Local Hilbert Subspaces

Assuming Bob reads 1 ($\lambda = 1$). During the entire measurement process, Bob experiences three stages (before, during, and after the measurement), as described by Eqs (11-12):

$$|\Psi\rangle \otimes |\,\text{ready}\,\rangle_O\,|E\rangle_L \rightarrow \alpha_0\,|0_B\rangle + \alpha_1\,|1_B\rangle\,|\,\text{reads } 1\rangle_O \rightarrow |\Psi_B\rangle \otimes |\,\text{ready}\,\rangle_O \qquad (23)$$

The branched local Hilbert subspaces are eventually relocated into the environment at large by unitary transformations, complying with the No-Hiding Theorem [23]:

$$U_E : |\Psi_B\rangle \otimes |E\,\rangle \rightarrow |E'\,\rangle \qquad (24)$$

**2.3. The Observer's Local View of the Measurement:**

In quantum measurements or quantum computing, the observer must repeatedly measure the same initial states. Each time, he reads one possible outcome, with the probability predicted by the Born rule, which leads to the following density matrix [33, p.53]:

$$\rho = \sum_{k=1}^{D} |\,g_k\rangle\,|\,c_k\,|^2\,\langle\, g_k\,|, \quad \sum_{k=1}^{D} |\,c_k\,|^2 = 1 \qquad (25)$$

Locally, the observer sees that the initial pure state, Eq. (1), with zero von Neumann entropy [33, p.179], becomes a mixed state, and its von Neumann entropy is increased to:

$$S(\rho) = -\mathrm{Tr}(\rho \ln \rho) = -\sum_{k=1}^{D} \{|\,c_k\,|^2 \ln |\,c_k\,|^2\} > 0 \qquad (26)$$

The observer concludes that his measurement is irreversible because the system's entropy increases and certain information is lost. However, in the entire Hilbert space encompassing all branches, there is no loss of information or gain in entropy. This is quite similar to the MWI, except that MWI comprises many independent, equally real worlds, whereas BHSI features numerous independent local Hilbert subspaces with predictable weights (probabilities).

## 3. Feature Comparison of CI, MWI, BM, and BHSI

| **Feature** | Copenhagen (CI) | Many-Worlds (MWI) | Bohmian Mechanics (BM) | BHSI |
|---|---|---|---|---|
| **1.** Wave Collapse? Unitarity? | Yes. Non-unitary | No. Fully unitary by splitting the global Hilbert space | No. Fully unitary (wavefunction guides particles) | No. Fully unitary by splitting the local Hilbert space |
| **2.** Ontology: Number of Worlds and "Me" | A single world, a single "Me." | Many real worlds, each with a "Me." | A single world, a single "Me." | A single world, a single "Me." |
| **3.** Probability: The Born Rule | Fundamental postulate (no deeper explanation) | Emergent from decision theory? (self-locating uncertainty?) | Explained by the equilibrium distributions of hidden variables | Interpreted as the weights of local Hilbert branches. |
| **4.** The Role of the Observer | Passive, external to the system, | Branching, then following one | Passive (particles have definite | Branching, engaging, then disengaging |
| | and causes collapse | world, and all worlds are real. | positions at all times) | from one Hilbert branch. |
| **5.** Determinism | Indeterministic (collapse introduces randomness) | Deterministic (but observers experience subjective randomness) | Deterministic (hidden variables define definite trajectories) | Deterministic (but observers experience local randomness) |
| **6.** Information Loss | Yes (collapse destroys superpositions permanently) | No (information persists in different worlds) | No (global wave function guides particles deterministically) | No (information persists in different Hilbert subspaces) |
| **7**. Can Branches Recombine? | N/A (only one world exists) | No (recoherence leads to identity crises) | N/A (only one world exists) | Yes? In theory, it is possible. |
| **8**. Locality of Physical Laws | Local (except for nonlocal collapse) | Local (no signal between branches) | Nonlocal (built into the global wave function) | No faster-than-light action |

Table 1. Feature Comparison of the Four Interpretations of Measurements

## 4. Comparing BHSI with MWI and CI by Examples

The BHSI is proposed as a "cost-effective" version of the MWI to avoid the collapse issue in the CI without the ontological excess of MWI. This section uses several examples to illustrate the similarities and differences between the three interpretations.

**Example 4.1**. The Double-Slit Experiment is the best-known experiment illustrating the *particle-wave duality* in quantum mechanics (QM) [15-16], using photons, electrons, and large C60 molecules [17]. When a particle hits the screen, the LHS in BHSI splits into an uncountable infinite number of branches (in theory), and the observer reads it at one position $x$.

$$|\Psi_B\rangle = \int dx'|x'\rangle\langle x'|\Psi_B\rangle[|\text{reads } x\rangle_O]^{\Delta(x,x')}, \quad \Delta(x,x') = \begin{cases} 1, & \text{if } x = x' \\ 0, & \text{if } x \neq x' \end{cases} \quad \text{(continuous case)} \tag{27}$$

$$|\langle x_B|\Psi_B\rangle|^2 = |\Psi(x)|^2 = |\Psi_I(x) + \Psi_{II}(x)|^2 \tag{28}$$

Because of the limitations of the experimental equipment, the integral in Eq. (27-28) should be replaced by a discrete summation over tiny pieces $\Delta_k$:

$$|\Psi_B\rangle = \sum_{k'} \Delta_{k'}|x_{k'}\rangle\langle x_{k'}|\Psi_B\rangle[|\text{reads } x_k\rangle_O]^{\Delta(k,k')}, \quad P(\Delta_k) = |\Psi_{\text{I}}(x_k) + \Psi_{\text{II}}(x_k)|^2\ \Delta_k \tag{29}$$

The BHSI and MWI rely on branching to maintain unitarity and interference without total information loss. In the BHSI, the observer disengages from the system after reading, and the interference or probability distribution (the Born rule) can be obtained naturally; however, in the MWI, the environment, coherent with each piece $\Delta_k$, is a whole world with a real observer. In a typical double-slit experiment, tens of thousands of photons hit the screen, and each photon updates thousands of branches. Because of the ontological issue, there is no convincing interpretation of probability in MWI yet: Many minds? Indexicalism? Decision theory? A rational bet on a particular result, or Envariance [8,9]?

The CI can explain the interference by simply assuming the Born rule. Still, each particle's hit causes a wave collapse (FTL action), thereby breaking unitarity and leading to information loss.

**Example 4.2**. *The Bell Tests of Entanglement*: Applying the Born rule, all three interpretations can explain the violation of the Bell inequality [18-19] without spooky actions at a distance between the paired particles or the two observers. However, the costs are different. In CI, the measurements made by Alice and Bob cause two wave collapses (FTL actions), resulting in information loss. MWI and BHSI have no collapse and no total information loss. But MWI ends with four Hilbert branches of worlds per photon pair, each containing Alice and Bob, while BHSI ends with four local Hilbert branches without multiple Alices and Bobs.

**MWI:** Alice and Bob update four worlds per photon pair, each containing an Alice and a Bob:

$$\mathrm{A}_1:|0\rangle_a\,|\mathrm{Alice}_{0,A}\rangle\,|\mathrm{Bob}_{0,A}\rangle\,|\mathrm{E}_{0,A}\rangle,\qquad \mathrm{A}_2:|1\rangle_a\,|\mathrm{Alice}_{1,\mathrm{A}}\rangle\,|\mathrm{Bob}_{1,\mathrm{A}}\rangle\,|\mathrm{E}_{1,\mathrm{A}}\rangle \tag{30}$$

$$\mathrm{B}_1:|0\rangle_b\,|\mathrm{Alice}_{0,B}\rangle\,|\mathrm{Bob}_{0,B}\rangle\,|\mathrm{E}_{0,B}\rangle,\qquad \mathrm{B}_2:|1\rangle_b\,|\mathrm{AliceE}_{1,\mathrm{B}}\rangle\mathrm{BobE}_{1,\mathrm{B}}\rangle\,|\mathrm{E}_{1,\mathrm{B}}\rangle \tag{31}$$

**BHSI**: Alice and Bob update four branches per photon pair in their local Hilbert space,

$$\mathrm{A}_1:|0_B\rangle_a(|\text{ reads }0\rangle_O)^{\Delta(\alpha,0)}\rightarrow|0_B\rangle_a,\qquad \mathrm{A}_2:|1_B\rangle_a(|\text{ reads }1\rangle_O)^{\Delta(\alpha,1)}\rightarrow|1_B\rangle_a,\quad \alpha\in\{0,1\} \tag{32}$$

$$\mathrm{B}_1:|0_B\rangle_b(|\text{ reads }0\rangle_O)^{\Delta(\beta,0)}\rightarrow|0_B\rangle_b,\qquad \mathrm{B}_2:|1_B\rangle_b(|\text{ reads }1\rangle_O)^{\Delta(\beta,1)}\rightarrow|1_B\rangle_b,\quad \beta\in\{0,1\} \tag{33}$$

Typically, millions of photon pairs are measured by Alice and Bob in a Bell test.

**Example 4.3**. *Wigner's Friend Thought Experiment* [5, 19-20] is a compelling example involving mixed observers. Setup: The Friend (F) observes a qubit state: $(|0\rangle + |1\rangle)/\sqrt{2}$ in a Lab; simultaneously, Wigner (W), outside, observes F and the qubit. What occurs?

**CI:** F collapses the qubit, and W sees what F sees. One collapse. Why? F is the preferred observer (he measures the qubit), and F is a classical object that cannot entangle with a qubit.

**MWI:** F updates two worlds in the global Hilbert space, each containing an F and a W:

$$\mathrm{H}_1:|0\rangle\,|\mathrm{F}_0\rangle\,|\mathrm{W}_0\rangle\,|\mathrm{E}_0\rangle,\qquad \mathrm{H}_2:|1\rangle\,|\mathrm{F}_1\rangle\,|\mathrm{W}_1\rangle\,|\mathrm{E}_1\rangle \tag{34}$$

At the same time, W also updates two worlds, each containing an F and a W, too:

$$\mathrm{H}_3:|0\rangle\,|\mathrm{F}_0\rangle\,|\mathrm{W}_0\rangle\,|\mathrm{E}_0\rangle,\qquad \mathrm{H}_4:|1\rangle\,|\mathrm{F}_1\rangle\,|\mathrm{W}_1\rangle\,|\mathrm{E}_1\rangle \tag{35}$$

There is no collapse, no preferred observer, and F can be entangled with a qubit. Moreover, we can set $H_1 = H_3$ and $H_2 = H_4$ because H1 & H3 (H2 & H4) are physically indistinguishable, leading to a single-branching, two-worlds scenario. No matter whether it is two or four worlds, there is no identity conflict. If F and W shake hands, they must see the same result and in the same world.

**BHSI**: Friend updates two decoherent local branches, engages one, and then disengages:

$$\mathrm{H}_1 : |0_B\rangle(|\text{reads }0\rangle_O)^{\Delta(\alpha,0)} \to |0_B\rangle, \qquad \mathrm{H}_2 : |1_B\rangle(|\text{reads }1\rangle_O)^{\Delta(\alpha,1)} \to |1_B\rangle, \quad \alpha \in \{0,1\} \tag{36}$$

Because Friend measures the qubit, his branching is dominant; the local Hilbert subspaces must be updated synchronously with his, so Wigner's two branches should synchronize with Friend's:

$$\mathrm{H}_3 : |0_B\rangle(|\text{F reads }0\rangle\,|\text{reads }0\rangle_O)^{\Delta(\alpha,0)} \to |0_B\rangle, \quad \mathrm{H}_4 : |1_B\rangle(|\text{F reads }1\rangle\,|\text{reads }1\rangle_O)^{\Delta(\alpha,1)} \to |1_B\rangle \tag{37}$$

Wigner will observe an outcome of 0 or 1 if his friend engages with H1 or $H_2$. Like the MWI, the process is unitary, with no information loss or collapse; the friend's state can be entangled with a qubit, with no preferred observer but a causally dominant branching. Similar to the CI, with only one world, one Wigner, and one Friend, they see the same result and can always shake hands. Suppose Alice and Bob are outside, watching Wigner or his friend simultaneously, and then they shake hands afterward. What happens? We will use this scenario in Appendix A.

**Example 4.4**. *The Black hole information paradox*: Hawking's semi-classical calculations suggest that black hole evaporation via Hawking radiation is thermal and random [22]. If so, it destroys information about the infalling matter, violating unitarity. MWI and BHSI have different branching structures (global vs. local) for modeling Hawking radiation, both of which are consistent with the No-Hiding theorem [23]. Still, MWI leads to many real worlds while BHSI leads to local branches. However, the Hawking radiation in the CI causes collapses and information loss, violating the No-Hiding theorem.

**Example 4.5**. The delayed choice quantum eraser experiments [24-25]: In the MWI and BHSI, all Hilbert branches were already recorded when signal photons (pair a) hit the screen. The observer later chooses which branches based on the path of the idle photons (pair b), i.e., which-way information ($w$) is kept ($w$ = 1, no interference) or erased ($w$ = 0, seeing interference). There is no collapse or retrocausality, but there are many worlds in MWI, compared to many local subspaces in BHSI. In the CI, reality is determined only when the measurement is complete, so retrocausality does not occur either. By the way, does the experiment imply retrocausality? No. The state of the entangled signal photon pair *a* and the idle photon pair *b* can be written as:

$$|\Psi\rangle = (1/\sqrt{2})\sum_{w=0}^{1} |w\rangle_a\,|w\rangle_b = (1/\sqrt{2})\left(|0\rangle_a\,|0\rangle_b + |1\rangle_a\,|1\rangle_b\right) \tag{38}$$

When photon pairs (a) hit the screen, they leave objective records for $w_a = 0$ and $w_a = 1$. Importantly, the interference ($w_a = 0$) can only be recovered if the timing-matched photon pairs (b) are later selected to take the path for $w_b = 0$, which occurs with probability $|\langle 00|\Psi\rangle|^2 = ½$. This resembles a Bell test: Alice's ability to interpret her data depends on receiving Bob's correlated records—regardless of their spatial distance from the source—but it requires no retrocausality, only pre-established entanglement.

## 5. Decoherence, Branch Independence and Possible Recoherence: MWI vs BHSI

### 5.1. The environment involved in MWI and BHSI

In MWI, each branch is a whole, independent, and real world. Within a world, "objects" have definite macroscopic states by fiat [Eq. (1), 8]:

$$|\Psi_{\text{WORLD}}\rangle = |\Psi_{\text{OBJ. }1}\rangle|\Psi_{\text{OBJ. }2}\rangle\cdots|\Psi_{\text{OBJ. }N}\rangle|\Phi\rangle \tag{39}$$

The product state is relevant only to variables used in the macroscopic description of the objects. There might be some entanglement between weakly coupled variables, which should belong to $|\Phi\rangle$. The universe is expressed as a superposition of all existing worlds:

$$|\Psi_{\text{UNIVERSE}}\rangle = \sum_i^M \alpha_i |\Psi_{\text{WORLD }i}\rangle, \qquad \sum_i^M |\alpha_i|^2 = 1 \tag{40}$$

Configuring a world with approximately $10^{80}$ particles is challenging (preferred basis?), and no one knows the total number *M*, except that it grows exponentially over time (just one double-slit experiment can branch into millions of worlds). As described in Case 3, Section 2.2, when measuring a qubit, one of the branches in Eq. (40) (where the observer lives) is entangled with the two-qubit states described in Eqs. (17) and (19), resulting in two independent worlds, each having a Bob. Although mathematically possible, recohering the two branches in Eq. (19) or any two in Eq. (40) is ontologically forbidden (it causes identity crises) and practically impossible on the scale of worlds.

Contrary to MWI, the branches in the BHSI are local Hilbert subspaces, and each observation triggers a branching in its own LHS. There is no need for a preferred basis: the basis chosen by the observer in Eq. (1) is the basis for branching. Based on the quantum decoherence theory [26-28], the branching operator in Eq. (3) can be understood as follows:

$$\hat{B}:\left(\sum_{k=1}^{N} c_k|g_k\rangle\right)\otimes|E\rangle_L \rightarrow \sum_{k=1}^{N} c_k|g_k\rangle|E_k\rangle_L \equiv \sum_{k=1}^{N} c_k|g_{B;k}\rangle, \quad {}_L\langle E_i|E_k\rangle_L \approx \delta_{i,k} \tag{41}$$

Here, $|E\rangle_L$ represents the minimal local environment, which directly interacts with the quantum system and contains about 10 ~100 particles. The nature of branching is the same for MWI and BHSI. The difference lies in the size of their respective environments: a whole world versus the local environment (maximal versus minimal, or $10^{80}$ versus $10^2$). Therefore, controlled recoherence in BHSI is both mathematically and ontologically permissible and practically conceivable. If the environment is fully controlled, one can construct a debranching operator for the recoherence of decohered branches, following Eq. (9-10) and assuming the time interval between decoherence and recoherence is $\tau$:

$$B^\dagger(\alpha_1 e^{i\Phi_1(\tau)}|\psi_1\rangle|E_1\rangle_L + \alpha_2 e^{i\Phi_2(\tau)}|\psi_2\rangle|E_2\rangle_L) = e^{i\Phi_1(\tau)}(\alpha_1|\psi_1\rangle + \alpha_2 e^{i\Delta\Phi_2(\tau)}|\psi_2\rangle)\otimes|E\rangle_L \tag{42}$$

Note that the accumulated relative *phase shift* $\Delta\Phi$ vanishes if there is no branch-dependent quantum interaction with their environments. In the next sections, the quantum teleportation [29] and the proposed experiments using full-loop Stern-Gerlach Interferometers (SGI) [30-31] demonstrate locally controlled decoherence-recoherence processes (CDRP) in BHSI.

### 5.2. Teleportation: decoherent subspaces and the recoherence process in BHSI

Assume that Alice has a pair of photons *C* and *D*, entangled in the Bell state $|B_1\rangle_{CD} = |\Phi^+\rangle_{CD}$, and she also has a photon *B* rotated to the following state:

$$|\psi\rangle_B = \alpha\,|0\rangle_B + \beta\,|1\rangle_B\,,\ |\alpha|^2 + |\beta|^2 = 1 \tag{43}$$

Photon *D* will teleport this state [33]. Before the swap, Alice has three photons (C, B, and D), while Bob receives one (*D*). The state of the three photons is given by a separable pure state in the 8-dimensional product Hilbert space:

$$|\Psi\rangle = |\Phi^+\rangle_{CD} \otimes |\psi\rangle_B = \frac{1}{\sqrt{2}}\left(|0\rangle_C \otimes |0\rangle_D + |1\rangle_C \otimes |1\rangle_D\right) \otimes \left(\alpha\,|0\rangle_B + \beta\,|1\rangle_B\right) \tag{44}$$

Then, photons *C* and *B* are entangled to the four Bell states ($B_k$) by unitary swapping $U_A$, forcing photon *D* to carry correspondingly rotated states from photon *B* [33. P.165]:

$$U_A:\ \ |\Psi\rangle = \frac{1}{2}\Big[|\Phi^+\rangle_{CB} \otimes (\alpha\,|0\rangle_D + \beta\,|1\rangle_D) + |\Phi^-\rangle_{CB} \otimes (\alpha\,|0\rangle_D - \beta\,|1\rangle_D)$$

$$+\,|\Psi^+\rangle_{CB} \otimes (\beta\,|0\rangle_D + \alpha\,|1\rangle_D) + |\Psi^-\rangle_{CB} \otimes (\beta\,|0\rangle_D - \alpha\,|1\rangle_D)\Big] \equiv \sum_{k=1}^{4} |B_k\rangle_{CB}\,|\psi_k\rangle_D \tag{45}$$

We can rewrite Eq. (45) as the decoherent state in Eq. (43) of photon *B,* realized in photon *D*:

$$U_A:\ \ |\Psi\rangle = \alpha\,|0\rangle_D \otimes |E_0\rangle_L + \beta\,|1\rangle_D \otimes |E_1\rangle_L\,,\ \ {}_L\langle E_0\,|\,E_1\rangle_L = 0 \tag{46}$$

$$|E_0\rangle_L \equiv (1/2)(|\Phi^+\rangle_{CB} + |\Phi^-\rangle_{CB}) + (\beta/2\alpha)\cdot(|\Psi^+\rangle_{CB} + |\Psi^+\rangle_{CB}) \tag{47}$$

$$|E_1\rangle_L \equiv (\alpha/2\beta)\cdot(|\Phi^+\rangle_{CB} - |\Phi^-\rangle_{CB}) + (1/2)(|\Psi^+\rangle_{CB} - |\Psi^+\rangle_{CB}) \tag{48}$$

Eq. (46) shows photon *D* is *locally decoherent with minimal environmental involvement,* an entangled photon pair *CB,* consistent with Eq. (22). Now, Alice chooses Bell states as her measurement basis, splitting the local system into four branches and forcing photon *D* to take one of the four possible states as described in Eq. (45). After receiving the record from Alice about which Bell state ($B_i$) she observes, Bob rotates photon *D* accordingly using a unitary transformation $U_i$, allowing him to fully recover the original state of photon *B* in the teleported photon *D*. Therefore, the operations on photon *D* can also be viewed as a local controlled decoherence-recoherence process (CDRP):

$$\alpha\,|0\rangle_B + \beta\,|1\rangle_B \xrightarrow{U_A} \alpha\,|0\rangle_D\,|E_0\rangle_{CB} + \beta\,|1\rangle_D\,|E_1\rangle_{CB} \xrightarrow{\hat{M}_i} |\psi_i\rangle_D \xrightarrow{U_i} \alpha\,|0\rangle_D + \beta\,|1\rangle_D \tag{49}$$

The teleportation process is not a true debranching process as described in Eq. (42): Bob's rotation $U_i$ does not merge the four branches of photons *BC* created by Alice's measurement; however, Eqs. (46–49) demonstrate that the local CDRP is *observable* in teleportation, not merely as a theoretical construct, and the locally decoherent subspaces can be recohered via unitary transformations ($U_i\hat{M}_i$) as prescribed in Eq. (42) with $\Delta\Phi = 0$.

Importantly, this interpretation is only suitable for BHSI, where the measurement-generated branches are also locally decoherent. Moreover, in MWI, any mismatch in messaging (e.g., the wrong message or a missed one) leads to ontological ambiguity: Bob cannot determine *which* Alice he inhabits a world with, since the four branches created by Alice's measurement are causally disconnected. BHSI avoids this ambiguity entirely—Alice and Bob inhabit a single world, and they can communicate classically and causally.

**5.3: Visualizing Controlled Decoherent Branches Using Stern-Gerlach Interferometers**

The BHSI makes two key predictions: decoherent branches evolve unitarily and independently while preserving their Born rule weights; quantum branching is reversible under unitary recoherence if the environment is fully controlled. We demonstrate that these predictions can be visualized and even directly measured using modern full-loop Stern-Gerlach Interferometers (SGIs [30,31]), offering empirical grounding for the BHSI's central features.

**5.3.1. Visualizing the controlled decoherence and recoherence process (CDRP)**: We propose the experiment to test CDRP using a single vertical full-loop SGI [30], where a test mass (with spin ½) is decohered and recohered through entanglement with spatial paths. The experiment (referring to the left arm below in Fig. 2 from [31]) involves three steps. The first step is *preparation*: the test mass $m_1$ (e.g., an atom or a nanodiamond with an NV center), located at $|C\rangle_1$, is initialized in a general spin-qubit state.

$$|\psi_0\rangle_1 = \sin\theta\,|\uparrow\rangle_1 + e^{i\phi}\cos\theta\,|\downarrow\rangle_1 \tag{50}$$

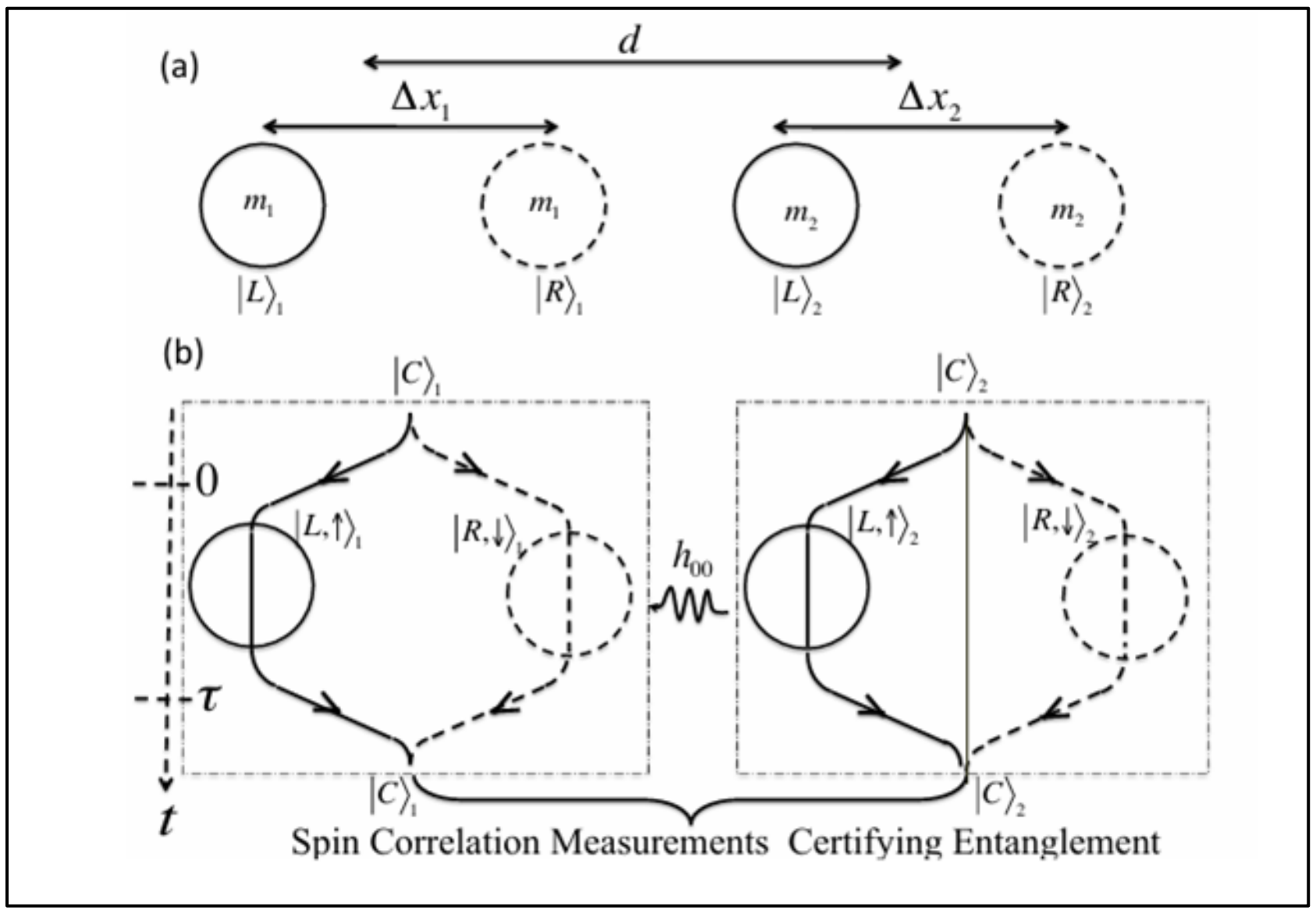

Fig 2: Using SGIs to test the local decoherence (from [31], Fig. 1)

This is achievable via rotated magnetic fields or RF pulses [30]. The second step is *decoherence*: dropping the test mass into the first vertical SGI, whose gradient magnetic field entangles spin with momentum, forming a superposition of two paths (left and right):

$$|\psi\rangle_1 |C\rangle_1 \to |\psi(0)\rangle_1 = \sin\theta |\uparrow, L\rangle_1 + e^{i\phi}\cos\theta |\downarrow, R\rangle_1 \tag{51}$$

This step simulates the BHSI "branching" process. Then it moves down through the space between the upper and lower SGIs for a certain duration. The third step is *recoherence*: the lower SGI's gradient magnetic field reverses the momentum splitting, ideally yielding:

$$|\psi(0)\rangle_1 \to |\psi(\tau)\rangle_1 = \sin\theta |\uparrow\rangle_1 + e^{i\phi}\cos\theta |\downarrow\rangle_1 \tag{52}$$

The spatial branches recombine unitarily, restoring the initial spin coherence. By measuring the two angles, one can observe the recoherence described by Eq. (42) with $\Delta\Phi = 0$.

**5.3.2. Visualizing the branch weight encoding the Born rule:** The faithful recovery of the initial θ angle in the CDRP, Eq. (52), indirectly verifies the branch weights. The fact that the probabilities $\sin^2\theta$ and $\cos^2\theta$ are maintained throughout the branching and recohering process indicates these weights are inherent to the branches. If the branches did not carry these amplitudes coherently, the original superposition could not be restored with the correct probabilities.

Furthermore, one can directly verify the branch weights encoding the Born rule. One can place a "which-way" path sensor (possibly charging the test mass) between the upper and lower SGIs to determine whether the test mass took the 'L' or 'R' path. Although such a measurement would inevitably destroy the interference pattern and prevent re-coherence, repeating the experiment many times would show that the measurement outcomes are statistically distributed:

$$P(\uparrow, L) = {}_1\langle\psi|\uparrow, L\rangle_1 = \sin^2\theta, \quad P(\downarrow, R) = {}_1\langle\psi|\downarrow, R\rangle_1 = \cos^2\theta \tag{53}$$

This directly confirms that the decoherent branches carry the corresponding weights that preserve the Born rule, as prescribed by Eq. (15). Advances in high-resolution nanoparticle position sensors, capable of detecting the spatial location of individual test masses, are making such a which-way detection experimentally feasible for increasingly massive objects, especially charged masses.

**5.3.3. Visualizing the independent unitary evolution of the branches:** BHSI predicts that decoherent branches evolve independently but unitarily, enabling branch-dependent interactions to produce measurable phase shifts as shown in Eq. (42). Suppose the two branches of the left interferometer stay separated by a distance $\Delta x \sim 10$ μm for about $\tau \sim 100$ ms, and an interaction source ($m_2$) is located nearby at a distance $d \sim 100$ μm with potential $V(x)$ (see the two arms of Fig. 2). In this case, the accumulated phase shift for the left mass can be derived as follows:

$$|\psi_0\rangle_1 = \alpha|\uparrow\rangle_1 + \beta|\downarrow_1\rangle \to U(t)(\alpha|\uparrow, L\rangle_1 + \beta|\downarrow, R\rangle_1) \xrightarrow{t=\tau} e^{i\Phi_L(\tau)}[\alpha|\uparrow\rangle_1 + e^{i\Delta\Phi(\tau)}\beta|\downarrow\rangle_1]$$
$$\to \Delta\Phi(\tau) = \Phi_R(\tau) - \Phi_L(\tau) = [V(R) - V(L)]\tau/\hbar = [V(d - \Delta x_1/2) - V(d + \Delta x_1/2)]\tau/\hbar \tag{54}$$

**Testing gravitational phase shift**: The Bose *et al.* SGI- quantum gravity test ([31]) is designed to observe ΔΦ from a weak (Newtonian) gravitational coupling between two nearby masses ($m_1 \sim m_2 \sim 10^{-14}$ kg), as a branch-dependent interaction (see Fig. 2). To demonstrate the phase shift in a CDRP better, we turn off the gradient magnetic fields in the right arm so that the right test mass moves in a straight line in Fig. 2. If weak gravity is quantum and assume $m_1 \sim m_2 \sim 10^{-14}$ kg, $d \sim$ 100 μm, Δx ~ 10 μm, and τ ~ 100 ms, the phase shift is given by:

$$\Delta\Phi(\tau)=\frac{Gm_1m_2\tau}{\hbar}\left(\frac{1}{d-\Delta x_1/2}-\frac{1}{d+\Delta x_1/2}\right)\sim\frac{Gm_1m_2\tau\Delta x_1}{\hbar[d^2-(\Delta x_1)^2/4]}\sim 6\times10^{-3}\text{rad} \tag{55}$$

**Testing electromagnetic (EM) phase shift**: We can modify the above setting to test EM phase shift by charging the two test masses ($q_1 \sim q_2 \sim e$) while turning off the gradient magnetic fields in the right arm in Fig. 2. In this case, the phase shift is given by:

$$\Delta\Phi(\tau)=\frac{q_1q_2\tau}{\hbar}\left(\frac{1}{d-\Delta x_1/2}-\frac{1}{d+\Delta x_1/2}\right)\sim\frac{q_1q_2\tau\Delta x_1}{\hbar[d^2-(\Delta x_1)^2/4]}\sim\frac{q_1q_2\tau\Delta x_1}{\hbar d^2}\sim 0.2\text{ rad} \tag{56}$$

Here, we assume $q_1 \sim q_2 \sim -3e$, $d \sim$ 100 μm, Δx ~ 10 μm, and τ ~ 100 ms. Because the EM field is quantum and much stronger than the gravitational field, the EM phase shifts would be more significant and easier to detect by adjusting the charge, distance, and time parameters given above. This SGI test offers a perfect experimental confirmation of BHSI's independently and unitarily evolving branches in the CDRP as described in Eq. (42):

$$|\psi_0\rangle=(|\uparrow\rangle+|\downarrow\rangle)/\sqrt{2}\rightarrow U(t)(|\uparrow,L\rangle+|\downarrow,R\rangle)/\sqrt{2}\overset{t=\tau}{\rightarrow}|\psi_\tau\rangle=(|\uparrow\rangle+e^{i\Delta\Phi(\tau)}|\downarrow\rangle)/\sqrt{2} \tag{57}$$

Once again, these experimental interpretations are meaningful only within the BHSI framework, where measurement-induced decoherence is also local and potentially reversible. Therefore, to empirically differentiate BHSI from MWI, a definitive test must show that measurement-induced decoherence is either localized (i.e., it does not entangle the entire world) or conditionally reversible (i.e., allows for recoherence under controlled conditions). The relative phase shift ΔΦ may serve as a signature of certain recoherences. Whether it is possible to experimentally examine measurement-branched local Hilbert subspaces before they become irreversibly entangled with the environment remains an open challenge. Nonetheless, experiments such as delayed-choice and quantum eraser [24-25], modern SGIs [30-31], quantum error correction [32], or trapped-ion entanglement with photons [33] could be employed for this purpose.

## 6. Conclusion and Discussion

In the framework of the Branched Hilbert Subspace Interpretation (BHSI), a measurement is seen as a combination of unitary operators: branching, engaging, and disengaging. The branches are locally decoherent, evolving unitarily and independently, with their amplitudes determined by the initial system state. They may be conditionally remerged through recoherence. These features—locally decoherent but re-coherable branches—are key to BHSI's way of interpreting measurements, which differs from both the idea of permanently branched worlds and the concept of wavefunction collapse. Locally controlled decoherence-recoherence (CDRP) processes are observable in quantum teleportation. They can also be tested with modern Stern-Gerlach interferometers (SGIs), where CDRP, branch widths (which encode the Born rule), and branch-dependent electromagnetic (EM) phase shifts can be directly observed.

BHSI maintains the elegance of the Many-Worlds Interpretation (MWI), including its unitarity, information preservation, and collapse-free evolution, while avoiding its ontological excesses. It retains the simplicity of the Copenhagen Interpretation (CI), with its single world and observer, but removes its ad hoc collapse postulate. BHSI addresses the following issues:

- Ontological minimalism: No parallel worlds or the need for a preferred basis.
- Unitary preservation: No collapse, no information loss, nor nonlocal structures.
- Causal primacy: Branching is defined by decoherence dynamics, not observer choices.
- Born rule emergence: Probabilities arise from branch weights.
- Experimental illustration: As shown in teleportation and the proposed SGI-based tests.

Aside from CI, BHSI, and Bohmian Mechanics (BM), other Single-World Interpretations exist (see Appendix **A**). However, no experimental test has yet definitively distinguished any of them from MWI. Nevertheless, by invoking Occam's Razor [35], BHSI offers a minimalist middle ground between the CI's wavefunction collapse and MWI's branching universe.

The conceptual foundations and experimental implications of BHSI have been further elaborated in two companion works [36,37]. Conceptually, the Island of Coherence (IOC) is presented as the central physical entity—an operationally isolated, inseparable quantum system, described by a Local Hilbert Space (LHS), which possesses no spacetime metric, but coexists with background spacetime in which the IOC is embedded—a dual structure already implicit in first quantization. Upon measurement, this LHS dynamically splits into branched Hilbert subspaces. The Born rule and quantum nonlocality follow from the LHS inner-product structure [36]. Experimentally, a three-stage program using dual-sensing Stern–Gerlach interferometers is proposed to probe IOC lifecycles, fuzzy boundaries, and possible recoherence effects [37].

## Appendix A: Comparison of BHSI with Other Single-World Interpretations

This appendix briefly contrasts BHSI with three other single-world interpretations: QBism, Relational Quantum Mechanics (RQM), and Modal interpretations.

**BHSI versus QBism**: Objective Branching versus Subjective Belief. Both reject collapse and many-worlds, but diverge fundamentally in their ontologies. QBism treats quantum states as agent-centered beliefs, with probabilities as subjective Bayesian updates [38]. BHSI posits an agent-independent Hilbert space structure in which branching subspaces represent objective measurement records—physical correlates of decoherence. QBism addresses measurement through observer psychology; BHSI resolves it through unitary dynamics alone. In the Wigner's friend scenario (Example 4.1.3), all observers must align with the objectively existing branch defined by the lab equipment, regardless of personal beliefs.

**BHSI versus Relative QM** (RQM): Causal Structure versus Observer-Dependent Reality. Both reject absolute quantum states, but BHSI disagrees with RQM's claim that observables are inherently observer-dependent [39]. BHSI's branches develop deterministically via unitary evolution, with measurement results represented as objective subspace decompositions independent of observers. RQM faces conceptual instability in extended Wigner's friend scenarios (Example **4.3**): if facts are always relative, reconciling different observers' "facts" risks violating the causal structure: Who updates their state first? Alice, who observed the Friend? Bob, who observed Wigner? BHSI bypasses this: all observers align with the dominant causal branch defined by the system's objective decoherence history.

**BHSI versus Modal Interpretations**: Actual Branches versus "Possible" Properties. Both decompose Hilbert space into subsystems within a single world, but BHSI treats all branches as equally actual, not merely "possible" [40,41]. Modal interpretations rely on preferred factorizations (e.g., system–apparatus splits) to define provisional properties, leaving unclear when possibilities become facts. BHSI's branching is dynamically determined by unitary evolution—decoherence alone records outcomes without requiring contextual triggers. In Example **4.3**, all observers follow a shared, causally determined history; there is no "maybe."

In each case, BHSI eliminates ad hoc rules, psychological constructs, or arbitrary factorizations, relying solely on the geometry and dynamics of LHS to solve the measurement problem.

**Abbreviations**

| | |
|---|---|
| BHSI | Branched Hilbert Subspace Interpretation |
| BM | Bohmian Mechanics |
| CI | Copenhagen Interpretation |
| CDRP | Controlled Decoherence and Recoherence Process |
| EM | Electromagnetic |
| FTL | Faster Than Light |
| LHS | Local Hilbert Space |
| MWI | Many-Worlds Interpretation |
| QM | Quantum Mechanics |
| RQM | Relational Quantum Mechanics |
| SGI | Stern-Gerlach Interferometer |
| SWI | Single World Interpretation |